\documentclass[aps, pre,amsmath,amssymb,twocolumn,floatfix,superscriptaddress]{revtex4}

\newcounter{abc}
\renewcommand{\theequation}{\arabic{equation}\alph{abc}}

\usepackage{graphicx}
\usepackage{bm}
\usepackage{epsfig}

\newcommand{\be}{\begin{equation}}
\newcommand{\ee}{\end{equation}}
\newcommand{\bea}{\begin{eqnarray}}
\newcommand{\eea}{\end{eqnarray}}

\newcommand{\cl}{_{\rm cl}}

\newcommand{\ronesig}{\rho\one(\sigma)}
\newcommand{\rtwosig}{\rho\two(\sigma)}

\newcommand{\one}{^{(1)}}
\newcommand{\two}{^{(2)}}
\newcommand{\p}{^{(0)}}
\newcommand{\fex}{f^{\rm ex}}
\newcommand{\sig}{\sigma} %spheres diameter
\newcommand{\f}{f}
\newcommand{\fparent}{\f\p(\sigma)}
\newcommand{\parent}{\rho\p(\sigma)}

\begin{document}

\title{\bf Phase behaviour and particle-size cutoff effects in polydisperse fluids}

\author{Nigel B. Wilding}
\affiliation{Department of Physics, University of Bath, Bath BA2 7AY, United Kingdom}

\author{Peter Sollich}
\author{Moreno Fasolo}
\affiliation{Department of Mathematics, King's College London, Strand, London WC2R 2LS, UK}
\author{Matteo Buzzacchi}
\affiliation{Department of Physics, University of Bath, Bath BA2 7AY, United
Kingdom}

\date{\today}

\begin{abstract} 

We report a joint simulation and theoretical study of the liquid-vapor
phase behaviour of a fluid in which polydispersity in the
particle size couples
to the strength of the interparticle interactions. Attention is focussed
on the case in which the particles diameters are distributed according
to a fixed Schulz form with degree of polydispersity $\delta=14\%$. The
coexistence properties of this model are studied using grand canonical
ensemble Monte Carlo simulations and moment free energy calculations. We
obtain the cloud and shadow curves as well as the daughter phase density
distributions and fractional volumes along selected isothermal dilution
lines. In contrast to the case of size-{\em independent} interaction
strengths (N.B. Wilding, M. Fasolo and P. Sollich, J. Chem. Phys. {\bf
121}, 6887 (2004)), the cloud and shadow curves are found to be well
separated, with the critical point lying significantly below the cloud
curve maximum. For densities below the critical value, we observe that
the phase behaviour is highly sensitive to the choice of upper cutoff on
the particle size distribution. We elucidate the origins of this effect
in terms of extremely pronounced fractionation effects and discuss the
likely appearance of new phases in the limit of very large values of the
cutoff. 


\end{abstract} 
\maketitle
\setcounter{totalnumber}{10}

\section{Introduction and background}
\label{sec:intro}

When the constituent particles of a many body system exhibit variation
in terms of some attribute such as their size, shape or charge, then
that system is termed ``polydisperse''. Occurring, as it does, in soft
matter systems as diverse as colloids, polymers and liquid crystals, the
phenomenon of polydispersity is both widespread and basic. However,
owing to the complexity that polydispersity bestows on a system, gaining
an understanding of its physical consequences represents a considerable
challenge to theory, simulation and experiment alike. Meeting this
challenge is not solely a matter of fundamental concern, but also one of
practical importance: the presence of polydispersity in substances of
commercial and industrial importance (such as paints, fuels and
foodstuffs) is known to profoundly affect their thermodynamic and
processing properties in ways which are as yet neither well characterized
nor well understood \cite{LARSON99,FREDRICKSON,RUSSEL89}.  

Of the many fundamental questions that polydispersity elicits, one of
the most central concerns its influence on bulk phase separation. The
phase behavior of polydisperse fluids is known to be considerably richer
in both variety and character than that of their monodisperse
counterparts \cite{GUALTIERI82,SOLLICH02}. This richness has its source
in {\em fractionation} effects
\cite{EVANS98,EVANS01,HEUKELUM03,SHRESTH02,EDELMAN03,FAIRHURST04,ERNE05,WILDING04}: the
distribution of the polydisperse attribute differs from one coexisting
phase to another. In order to quantify this effect, it is expedient
\cite{SALACUSE} to regard the polydisperse attribute as a continuous
variable $\sigma$, and to define a {\em density} distribution
$\rho(\sigma)$, where $\rho(\sigma)d\sigma$ is the number density of
particles in the range $\sigma\ldots\sigma+d\sigma$. Fractionation then
occurs when, at two phase coexistence, a ``parent'' density distribution
$\rho^{(0)}(\sigma)$ splits into two distinct ``daughter'' phase
distributions $\rho^{(1)}(\sigma)$ and $\rho^{(2)}(\sigma)$. Particle
conservation implies that the daughter distributions are related to the
parent via a simple volumetric average:

\be
(1-\xi)\rho^{(1)}(\sigma)+\xi\rho^{(2)}(\sigma)=\rho^{(0)}(\sigma)\:,
\label{eq:lever}
\ee
where $1-\xi$ and $\xi$ are the fractional volumes of the two phases.

Experimentally, the situation typically considered (for example in
studies of colloidal dispersions) is the nature of the phase behaviour
along a so-called {\em dilution line}. Here, one takes a sample of some
prescribed polydispersity and observes its coexistence properties as it
is progressively diluted by adding solvent, the temperature being held
constant. Under such circumstances the parent distribution maintains a
fixed shape, and only its overall scale varies according to the degree
of dilution. Accordingly, one can write

\be 
\parent=n\p\fparent\:, 
\ee 
where $\fparent$ is a fixed shape function which serves to define the form of the
polydispersity, while $n\p$ is the overall (parent) number density of
the system, whose value parameterizes the location of
the system on the dilution line.

\begin{figure}
\includegraphics[type=eps,ext=.eps,read=.eps,width=7.0cm]{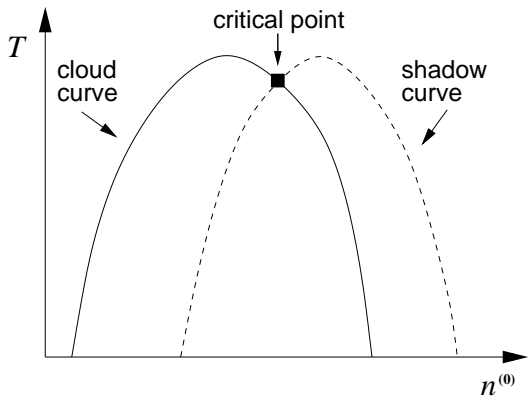}
\caption{A schematic fluid-fluid phase diagram for a polydisperse
fluid, showing temperature $T$ against density; the cloud curve gives
the parent density $n\p$ where phase separation first occurs while the
shadow curve records the density of the incipient coexisting phases.
Note, however, that the shadow phase density distribution lies off the
dilution line.}
\label{fig:phasediagram}
\end{figure}

In order to construct the full phase diagram for such a system one must
obtain the dilution line phase behaviour for the range of temperatures
of interest. However, in contrast to a monodisperse system where the
limits of phase stability and the densities of the coexisting phases are
completely specified by the coexistence binodal in the $n\p-T$ plane,
fractionation implies that polydispersity splits the binodal into {\em
cloud} and {\em shadow} curves \cite{SOLLICH02}, as shown schematically
in fig.~\ref{fig:phasediagram}. These curves mark, respectively, the density of
the onset of phase separation and the density of the incipient (shadow)
daughter phase. For parent densities lying wholly within the coexistence
region, two daughter phases form, the properties of which vary
non-trivially with $n\p$. Hence a full specification of the coexistence
properties of a polydisperse system requires not only a determination of
the locus of the cloud curve, but also the dependence of $\xi$, $\ronesig$
and $\rtwosig$ on $n\p$ and $T$.

In the present work, we describe a joint simulation and theoretical study
of liquid-vapor phase separation for a model fluid in which
polydispersity affects not only the length scale but also the
strength of the interaction potential. Previously we have considered the
restricted case of size-independent interaction strengths, for which the
critical point is found to lie very close to the maximum of the cloud
curve \cite{WILDING04}. By contrast, for the present model it is
located substantially below the maximum (cf.
fig.~\ref{fig:phasediagram}), as is observed in many experiments on
complex fluids (see e.g.~\cite{DOBASHI}). One implication of this is
that phase coexistence occurs at and above the critical temperature,
provided the overall parent density is less than its critical value, and
we investigate this feature here. Another difference to the case of
size-independent interactions is an acute sensitivity of the
coexistence properties to the presence of rare large particles. We
study this phenomenon by varying a particle size cutoff parameter.

As regards previous simulation studies of polydisperse phase equilibria,
other authors have (to date) considered exclusively the case of variable
polydispersity, in which the shape of the parent distribution changes
according to the temperature and overall density
\cite{STAPLETON,BOLHUIS,KOFKE93,KRISTOF,PRONK04,BATES98}. As such, these
studies do not reflect the most commonly encountered experimental
situation in which the polydispersity of a system is fixed by the
process of its chemical synthesis. With the recent advent of bespoke
Monte Carlo (MC) simulation techniques \cite{WILDING03,WILDING04},
however,  studies of fixed polydispersity within the (appropriate) grand
or semi-grand canonical ensemble framework are now tractable. In the
present work we apply these methods to study the phase separation of a
fluid whose polydispersity assumes a prescribed functional
form.

A number of analytical studies of polydisperse phase equilibria have
also been reported in the literature. These typically seek to calculate
the system free energy as a function of a set of density variables.
Unfortunately, this task is complicated by the fact that the requisite
free energy $f[\rho\p(\sigma)]$ is a functional of $\rho\p(\sigma)$, and
therefore occupies an infinite dimensional space. For sufficiently
simple model free energies which generalize the van der Waals (vdW)
approximation, a direct attack on the solution of the phase equilibrium
conditions is nevertheless sometimes possible, see
e.g.~\cite{GUALTIERI82,XU00}. The reason for this is that such models
are normally ``truncatable''~\cite{SOLLICH01} so that the phase equilibrium conditions 
can be reduced to nonlinear equations for a finite number of variables.
This approach has been applied to the
study of phase separation in fluids exhibiting separate size and
interaction strength polydispersity, yielding predictions for the cloud
and shadow curves and critical parameters as a function of
polydispersity \cite{BELLIER00}. An alternative approach which more
systematically exploits the advantages of truncatable models is the
moment free energy (MFE) method \cite{SOLLICH01,SOLLICH02}. This
approximates the full free energy appropriately in terms of a ``moment
free energy'' which depends on a small number of density variables,
thereby permitting the efficient use of standard tangent plane
construction to locate phase boundaries. The MFE method delivers (for
the given model free energy) exact results for the location of critical
points and the cloud and shadow curves. With appropriate refinements,
it can also be used to obtain exact numerical solutions to the phase
coexistence conditions when two or more phases coexist in finite
amounts~\cite{ClaCueSeaSolSpe00,SOLLICH01,SpeSol02,SpeSol03a}.
The MFE method has been applied
to the study of phase behaviour in systems ranging from polydisperse
hard rods to the freezing of hard spheres
\cite{SOLLICH02,FASOLO03,KALY03,KALY04,RASCON03}, and we shall deploy it
again here to study the phase behaviour of the present model, based on
a refined van der Waals-type approximation to the excess free energy.

Our paper is organized as follows \cite{PRL}. In sec.~\ref{sec:methods}
we introduce our model, a polydisperse Lennard-Jones (LJ) fluid that we
study via grand canonical Monte Carlo (MC) simulation and the MFE
method. Additionally we sketch the MC simulation methodology and provide
some brief background to the MFE calculations.
Sec.~\ref{sec:pd} details our estimates for the cloud and shadow
curves, as well as the behaviour of the fractional volumes of the
daughter phases and their density distributions along selected
isothermal dilution lines in a range including the critical
temperature. We then
turn in sec.~\ref{sec:cutoff} to a consideration of the effects of the
choice of the upper cutoff on the particle size distribution. Our
results demonstrate that coexistence properties can be acutely sensitive
to this choice. We elucidate the origin of this sensitivity in terms of
the interplay between the rate of decay of the parent size distribution at
large $\sigma$, and the $\sigma$-dependence of the excess chemical
potential. Likely implications for the phase behaviour at very large
cutoffs are also discussed. Finally, Sec.~\ref{sec:concs} details out
conclusions and highlights some outstanding questions worthy of further
investigation. 

\section{Model and methodologies}
\label{sec:methods}

\subsection{Simulation model}

We consider a polydisperse Lennard-Jones fluid whose interparticle potential takes the form:

\be
u_{ij}=\epsilon_{ij}\left[\left({\sigma_{ij}}/{r_{ij}}\right)^{12}
-\left({\sigma_{ij}}/{r_{ij}}\right)^{6}\right]
\label{eq:uij}
\ee
with interaction strength $\epsilon_{ij}=\sigma_i\sigma_j$ and
interaction radius $\sigma_{ij}=(\sigma_i+\sigma_j)/2$;
$r_{ij}=|{\bf r}_i-{\bf r}_j|$ denotes the separation between the particles.
A truncation was applied to the potential for
$r_{ij}>2.5\sigma_{ij}$ and no tail corrections were used.

With regard to the form of this potential, a couple of remarks are
appropriate.  Firstly, although it differs from that studied in
Ref.~\cite{WILDING04} only in the $\sigma$-dependence of the interaction
strength 
parameter $\epsilon_{ij}$, we shall show that this difference is crucial
in determining many of the qualitative aspects of the phase behavior of
the model. Secondly, the form of Eq.~(\ref{eq:uij}) might, in one sense,
be regarded as somewhat artificial because for
vanishingly small particles the interaction strength approaches zero.
An arguably more realistic form would be a repulsive hard core plus an
attractive well of variable strength.
However, it transpires that for the form of parent
distribution studied in the present work this drawback is of minor
significance since the physics of the system is dominated by the largest
particles.

\subsection{Parent distribution}
\label{sec:parent}

\begin{figure}
\includegraphics[type=eps,ext=.eps,read=.eps,width=7.0cm,clip=true]{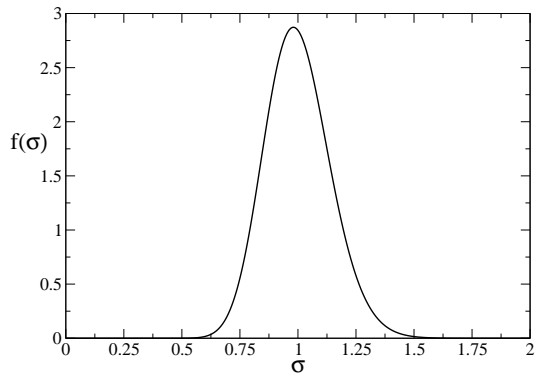}
\caption{Schulz distribution for the case $z=50$ (cf. Eq.~(\ref{eq:schulz})).}
\label{fig:schulz50}
\end{figure}

Throughout this work we consider the case in which the particle diameters $\sigma$ are drawn from a
(parental) distribution of the Schulz form \cite{SCHULZ}:
\be
f\p(\sigma)=\frac{1}{z!}\left(\frac{z+1}{\bar{\sigma}}\right)^{z+1}\sigma^z\exp\left[-\left(\frac{z+1}{\bar{\sigma}}\right)\sigma\right]\:,
\label{eq:schulz}
\ee
with a mean diameter $\bar{\sigma}$ which sets our unit length scale. We
have elected to study the case $z=50$, corresponding to a moderate degree of polydispersity: the
standard deviation of particle sizes is $\delta\equiv
1/\sqrt{z+1}\approx 14\%$ of the mean. The form of the distribution is
shown in Fig.~\ref{fig:schulz50}. Although our motivation for employing
the Schulz distribution is primarily ad-hoc, we note that it has been
found to fairly accurately describe the polydispersity of some polymeric
systems \cite{MCDONNEL}.

In both the simulations and the MFE calculations described below, the
distribution $f\p(\sigma)$ was limited to within the range
$0.5<\sigma<\sigma_c$. The upper cutoff $\sigma_c$ serves to prevent the
appearance of arbitrarily large particles in the simulation, but would
also be expected in experiment because in the chemical synthesis of
colloid particles, time or solute limits restrict the maximum particle
size \cite{KVITEK05}. Various choices of $\sigma_c$ have been
considered, and these are discussed in relation to the results of
Secs.~\ref{sec:pd} and \ref{sec:cutoff}.

\subsection{Simulation strategy}

The phase behaviour of the polydisperse LJ model Eq.~\ref{eq:uij}) was
studied via MC simulations within the grand canonical ensemble (GCE).
The MC algorithm invokes four types of operation: particle
displacements, deletions, insertions, and resizing. The particle
diameter $\sigma$ is treated as a continuous variable in the range $0\le
\sigma \le \sigma_c$, with $\sigma_c$ the prescribed cutoff. However,
distributions defined on $\sigma$ such as the instantaneous density
$\rho(\sigma)$, and the chemical potential $\mu(\sigma)$, are
represented as histograms defined by discretising the permitted range of
$\sigma$ into $120$ bins.  Most of the simulation results presented
below pertain to a periodic cubic system of linear dimension
$L=15\bar\sigma$, although near the critical point, where finite-size
effects are important, we determined the phase behaviour using system
sizes ranging up to $L=30\bar\sigma$. For further details concerning the
simulation algorithm, as well as the structure, storage and acquisition
of data, we refer the interested reader to ref.~\cite{WILDING02}. 

The principal observable of interest is the fluctuating form of the
instantaneous density distribution $\rho(\sigma)$.  From this we derive the
distribution $p(n)$ of the overall number density $n=\int
d\sigma \rho(\sigma)$, and that of the volume fraction
$\eta=(\pi/6)\int d\sigma \sigma^3\rho(\sigma)$. The existence of phase
coexistence at given chemical potentials is signalled by the presence of
two distinct peaks in the probability distribution $p(n)$. In order to
obtain estimates of dilution line coexistence properties at some
prescribed temperature, we employ an approach recently proposed by
ourselves in Ref. \cite{BUZZACCHI06}. For a given choice of $n\p$, the
method entails tuning the chemical potential distribution $\mu(\sigma)$
together with a parameter $\xi$, such as to simultaneously satisfy both
a generalized lever rule {\em and} an equal peak weight criterion
\cite{BORGS} for $p(n)$:

\setcounter{abc}{1}
\bea 
\label{eq:methoda}
n\p\f\p(\sigma) &=& (1-\xi)\ronesig + \xi\rtwosig \\
\addtocounter{abc}{1}
\addtocounter{equation}{-1}
r&=&1
\label{eq:methodb}
\eea 
\setcounter{abc}{0}
Here the daughter density distributions $\ronesig$ and $\rtwosig$ are assigned
by averaging only over configurations belonging to either peak of
$p(n)$. The quantity $r$ is the peak weight ratio: 
\be
r=\frac{\int_{n>n^{*}}p(n)dn}{\int_{n<n^{*}}p(n)dn}\:,
\ee
with $n^{*}$ a convenient threshold density intermediate between vapor
and liquid densities, which we take to be the location of the minimum
in $p(n)$. The tuning of $\mu(\sigma)$ and $\xi$ necessary to simultaneously
satisfy Eqs.~(\ref{eq:methoda}) and (\ref{eq:methodb}) can be
efficiently achieved by a combination of histogram extrapolation
techniques \cite{HR}, and a non-equilibrium potential refinement (NEPR)  procedure as described elsewhere \cite{WILDING03}. 

The value of $\xi$ resulting from the application of the above procedure
is the desired fractional volume of the liquid phase at the nominated
value of $n\p$. Cloud points are determined as the value of $n\p$ at
which $\xi$ first reaches zero (vapor branch) or unity (liquid branch),
while shadow points are given by the density of the coexisting
incipient daughter phase, which may be simply read off from the
appropriate peak density in the cloud point form of $p(n)$. It should be
pointed out that the finite-size corrections to estimates of coexistence
properties obtained using the equal peak weight criterion for $p(n)$ are
exponentially small in the system size \cite{BUZZACCHI06}. 

In order to obtain the phase behaviour of our model system, we scanned
the dilution line for a selection of fixed temperatures. We started by
setting $T=T_c$, the critical temperature, and tracked the locus of the
dilution line in a stepwise fashion. This tracking procedure must be
bootstrapped with knowledge of the form of $\mu(\sigma)$ at some initial
point on the dilution line.  A suitable estimate was obtained, for
a point near the critical density, by means of the NEPR procedure
\cite{WILDING03}, in conjunction with the equal peak weight criterion
for $p(n)$, discussed above.  Simulation data accumulated at this
near-critical state point was then extrapolated to a lower, but nearby
density $n\p$ by means of histogram reweighting, thus providing an
estimate of the corresponding form of $\mu(\sigma)$. The latter was
employed in a new simulation, the results of which were similarly
extrapolated to a still lower value of $n\p$. Iterating this procedure
thus enabled the systematic tracking of the whole dilution line.
Histogram extrapolation further permitted a determination of dilution
line properties at adjacent temperatures, thereby facilitating a
systematic determination of the phase behaviour in the $n\p-T$ plane.
Implementation of multicanonical preweighting techniques \cite{BERG92}
at each coexistence state point ensured adequate sampling of the
coexisting phases in cases where they are separated by a large
interfacial free energy barrier (see refs.~\cite{WILDING04,WILDING01}
for a fuller account of this latter procedure). 

\subsection{Moment free energy method}
\label{sec:mfe}

We complement the simulations with theoretical phase behaviour
calculations, following closely our study of the purely
size-polydisperse case~\cite{WILDING04}. 
To find a suitable expression for the excess free energy density
$\fex$, we approximate the {\em repulsive} part of our LJ interaction as
completely hard. The resulting contribution to $\fex$ is represented
accurately by the BMCSL approximation $\fex_{\rm
BMCSL}$~\cite{BOUBLIK,MANSOORI,SALACUSE}:
\[
\frac{\pi}{6}\beta\fex_{\rm BMCSL} = \left(\frac{\rho_2^3}{\rho_3^2} -
\rho_0\right)\ln(1-\rho_3) + 
\frac{3 \rho_1 \rho_2}{1-\rho_3} + \frac{\rho_2^3}{\rho_3(1-\rho_3)^2}\;.
\label{eq:BMCSL}
\]
This is a
function of the moments up to third order of the density distribution,
$\rho_i = (\pi/6)\int\! d\sigma\, \sigma^i\rho(\sigma)$ ($i=0,\ldots,
3$); note that $\rho_0=(\pi/6)n$ is proportional to the overall number
density, while $\rho_3=\eta$ is the volume fraction. We have set
$k_{\rm B}=1$ and defined $\beta=1/T$ as usual so that $\beta\fex$ has
dimension of density.

We treat the remaining {\em attractive} part of the LJ interaction
potential in the 
simplest possible way, by adding a quadratic van der Waals term to
the excess free energy. Using the fact that the interaction volume of two
particles of 
diameters $\sigma$ and $\sigma'$ scales as $(\sigma+\sigma')^3$, while
the interaction strength $\epsilon_{ij}$ is proportional to
$\sigma\sigma'$, this can be written as
%%%%%%%%%%%%%%
\begin{eqnarray}
\frac{\pi}{6}\beta\fex_{\rm vdW} &=& 
- \frac{1}{2t} \left(\frac{\pi}{6}\right)^2\int\! d\sigma\, d\sigma'\, 
\rho(\sigma) \rho(\sigma') (\sigma\sigma')(\sigma+\sigma')^3
\nonumber
\\
&=& - \frac{1}{t}(\rho_1 \rho_4 + 3\rho_2 \rho_3)
\label{fex}
\end{eqnarray}
%%%%%%%%%%%%
where $t$ is an appropriate dimensionless temperature.
Given the approximate character of the overall excess free energy
$\fex=\fex_{\rm BMCSL}+\fex_{\rm vdW}$, it would not
make sense to try to scale $t$ precisely to
the temperature in the simulations. Instead, we will be content to
study whether
our $\fex$ can reproduce the qualitative trends observed in the
simulations. While still somewhat crude, it should be better suited
to this task than previous versions~\cite{BELLIER00} because it incorporates
polydispersity not only into the attractive contribution, but also
into the hard core reference system.

As pointed out in the introduction, to predict phase equilibria for
polydisperse systems is generally 
still a computationally demanding task even if an explicit expression
for the excess free energy is available~\cite{SOLLICH02}.  The above
excess free energy has the simplifying feature that it only depends on
the finite set of moments $\rho_0,\ldots,\rho_4$, i.e.\ it is
truncatable~\cite{SOLLICH01}. This allows us to exploit the MFE method
to obtain accurate numerical predictions for
the phase behaviour. We refer the interested
reader to our analogous study of
the purely size-polydisperse case for a fuller description of the
methodology~\cite{WILDING04}.

\section{Phase behaviour: general aspects}
\label{sec:pd}

\begin{figure}[h]
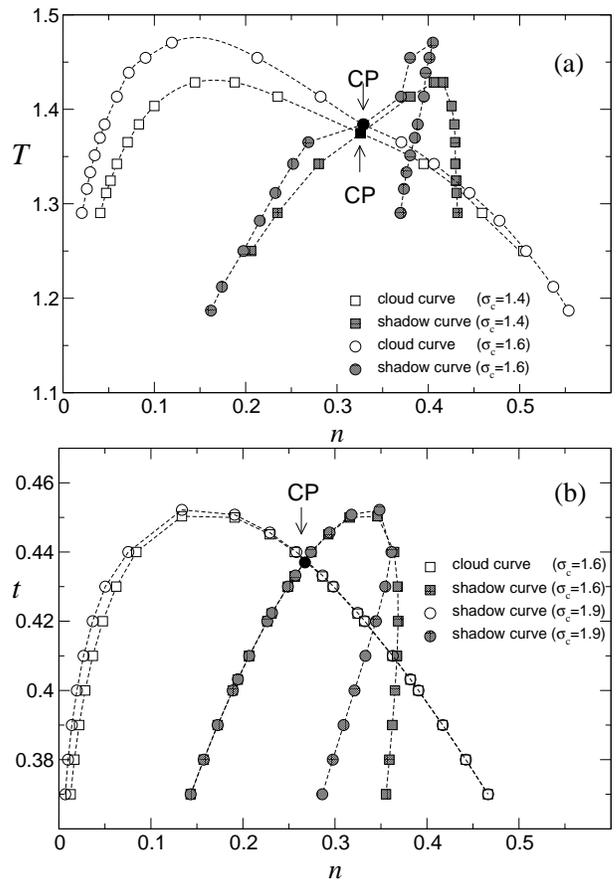

\includegraphics[type=eps,ext=.eps,read=.eps,width=8.0cm,clip=true]{Figs/Fig03a}
\includegraphics[type=eps,ext=.eps,read=.eps,width=8.0cm,clip=true]{Figs/Fig03b}
\caption{Cloud and shadow curves in the $n-T$  plane. {\bf (a)} MC
simulation results. {\bf (b)} MFE predictions. Lines are a guide to the eye.}
\label{fig:cl_sh_n}
\end{figure}

We have employed the simulation and MFE methods outlined in the previous
section to determine the cloud and shadow curves for our system. Various
choices of the particle size cutoff have been considered, although the
majority of our simulation results pertain to the cases $\sigma_c=1.4$
and $\sigma_c=1.6$. Referring to Fig.~\ref{fig:schulz50} it is clear
that both values are quite far out into the tail of the parent size
distribution, and naively one would therefore expect only very minor
differences. Our results, expressed in the $n$-$T$ representation, are
shown in Fig.~\ref{fig:cl_sh_n}(a) (for simulations of the LJ model) and
Fig.~\ref{fig:cl_sh_n}(b) (for MFE predictions). In both cases we
observe a strong separation of cloud and shadow curves; as a consequence
the critical point lies well below the cloud curve maximum. We also note
a pronounced difference in the cloud curves  for the two cutoffs,
contrary to naive expectation; detailed discussion of  this effect is
deferred to Sec.~\ref{sec:cutoff}.

Finite-size scaling methods \cite{WILDING95} (specifically the
matching of $p(n)$ to its known universal critical point form) were
utilized to provide accurate estimates of the critical parameters of the
LJ model. These yielded $n\p_c=0.326(3)$, $T_c=1.375(2)$ (for
$\sigma_c=1.4$);  $n\p_c=0.329(3)$, $T_c=1.384(2)$ 
(for $\sigma_c=1.6$),
which are to be compared with those of the monodisperse LJ fluid
\cite{WILDING95}: $n\p_c=0.3196(4)$, $T_c=1.1876(3)$. Thus polydispersity
of the form considered here acts to raise the critical
temperature substantially, although the associated increase in the
critical density is much more modest. We note that a
polydispersity-induced increase in $T_c$  has also been observed for the
case of size-independent interaction strengths \cite{WILDING04}. There,
however, the magnitude of the shift in $T_c$ was only around $5\%$ of
that seen in the present model, for the same parent distribution. 

The MFE results of Fig.~\ref{fig:cl_sh_n}(b) show qualitatively very
similar behaviour to the simulation results: the high density part of
the shadow curve has an unusual positive slope which becomes more
pronounced as the cutoff increases; at the same time the maxima of
cloud and shadow curves shift to higher temperatures. Quantitatively
the cutoff effects are somewhat weaker than in the simulations, which
is why we chose to display results for $\sigma_c=1.6$ and 1.9 rather than 1.4
and 1.6 as for the simulations.

One difference between simulations and theory is
the nature of the cloud and shadow curves in the vicinity of the 
critical point. In the simulation results ``dips'' are observable in
this region, in contrast to the MFE results which show no such effect.
The precise origin
of the dips is not clear to us at present. One possibility is that they
are a finite-size effect caused by the breakdown near the critical point
of our procedure for determining cloud points. Alternatively they may be
indicative of genuine critical point singularities in the cloud and shadow
curves which are not picked up by the theory because our approximate
free energy expression is of a mean-field type.

\begin{figure}[h]
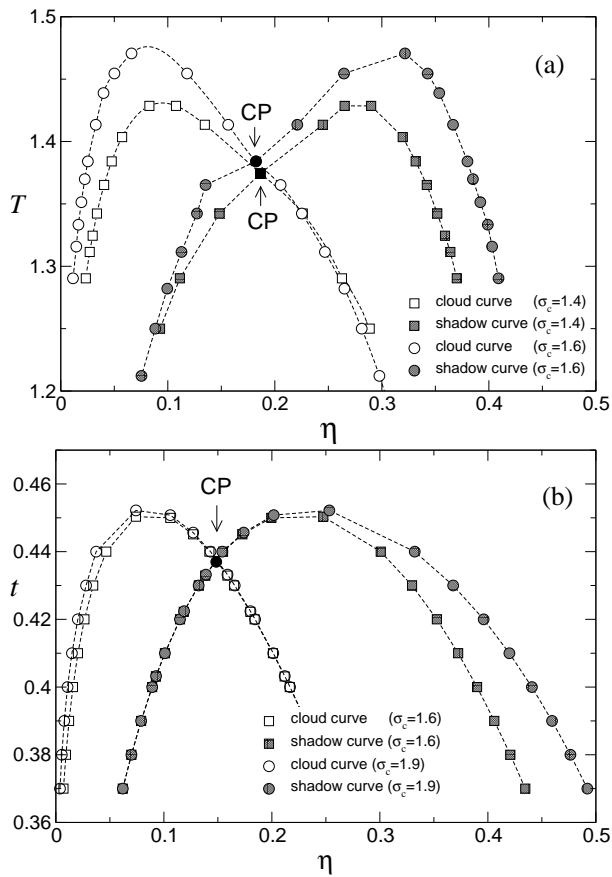

\includegraphics[type=eps,ext=.eps,read=.eps,width=8.0cm,clip=true]{Figs/Fig04a}
\includegraphics[type=eps,ext=.eps,read=.eps,width=8.0cm,clip=true]{Figs/Fig04b}
\caption{ Cloud and shadow curves in the $\eta-T$  plane. {\bf (a)} MC
results. 
{\bf (b)} MFE predictions. Lines are a guide to the eye.}
\label{fig:cl_sh_eta}
\end{figure}

In Fig.~\ref{fig:cl_sh_eta} we show the forms of the cloud and shadow
curves in the $\eta$-$T$ representation, i.e.\ plotting on the
$x$-axis the volume fraction $\eta$ rather than the density $n$ of the
cloud and shadow phases. Since the cloud phase always has the fixed
parental size distribution, the cloud curve is simply rescaled by this
change of representation; the same is not true of the shadow curve, however,
since the size distribution in the shadow varies throughout the phase
diagram. As a consequence, we observe that in the volume fraction plot
the cloud and shadow curves
separate even further than before, forming a rather symmetrical
``butterfly'' shape.
Again there is good qualitative agreement with the MFE predictions.

\begin{figure}[h]
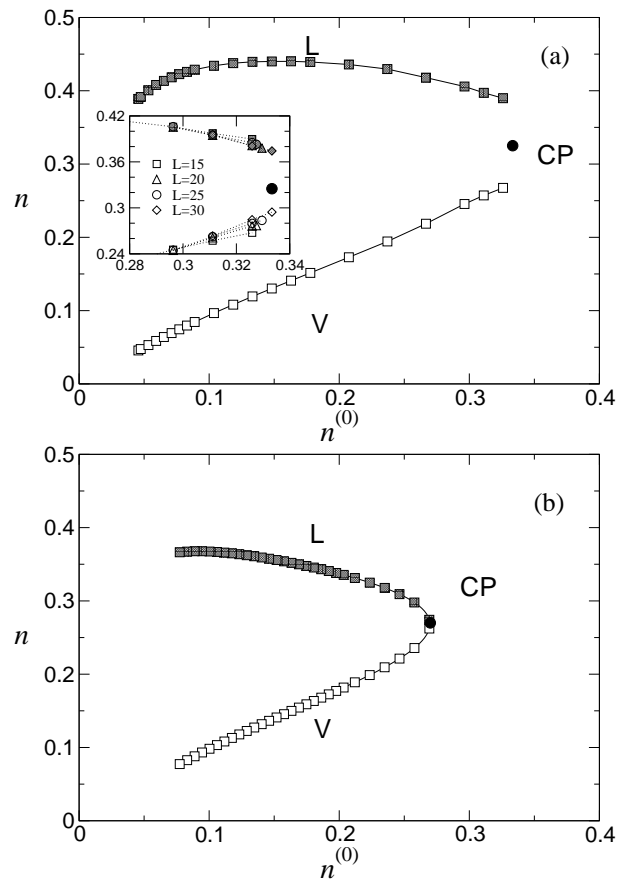

\includegraphics[type=eps,ext=.eps,read=.eps,width=8.0cm,clip=true]{Figs/Fig05a}
\includegraphics[type=eps,ext=.eps,read=.eps,width=8.0cm,clip=true]{Figs/Fig05b}
\caption{ {\bf (a)} MC results for the number densities of the coexisting vapor and liquid
phases as a function of $n\p$ at $T=T_c$, for $\sigma_c=1.6$. The inset
expands the near-critical region and shows data for a number of system sizes
(expressed in units of $\bar\sigma$),
as indicated. {\bf (b)} MFE predictions.}
\label{fig:den_n0}
\end{figure}

\begin{figure}[h]
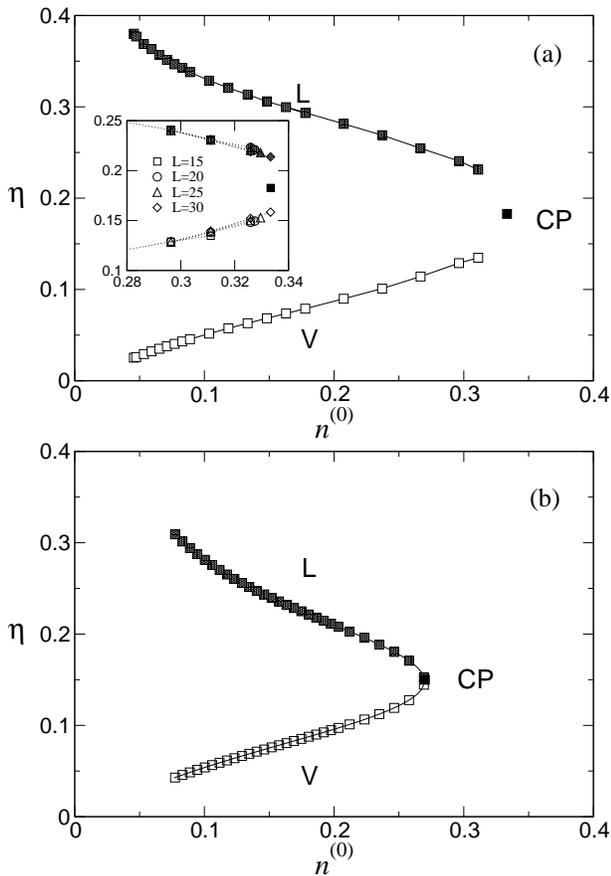

\includegraphics[type=eps,ext=.eps,read=.eps,width=8.0cm,clip=true]{Figs/Fig06a}
\includegraphics[type=eps,ext=.eps,read=.eps,width=8.0cm,clip=true]{Figs/Fig06b}
\caption{{\bf (a)} MC results for the volume fraction of the coexisting vapor and liquid
phases as a function of $n\p$ at $T=T_c$, for $\sigma_c=1.6$. The inset
expands the near-critical region and shows data for a number of system sizes. {\bf (b)} MFE predictions.}
\label{fig:eta_n0}
\end{figure}

Turning now to the coexistence properties {\em inside} the coexistence
region, we plot, in Figs.~\ref{fig:den_n0} and \ref{fig:eta_n0}, the
measured $n\p$ dependence of the number densities and volume fractions
of the coexisting vapor and liquid phases on the critical isotherm.
In the MC simulations, these were obtained simply by reading off the
peak positions of $p(n)$
and $p(\eta)$ for the appropriate values of $n\p$ and $T$. Close to the
critical point, one expects large finite-size effects to occur due to
the divergent correlation length. These are indeed apparent in our
measurements of the peak densities of $p(n)$ and $p(\eta)$ for a number
of system sizes (see insets of Figs.~\ref{fig:den_n0}(a) and
\ref{fig:eta_n0}(a)). Specifically, since the form of $p(n)$ is doubly
peaked for finite-sized systems, even at the critical point
\cite{WILDING95}, one expects that the peak densities approach the
critical density as a universal power law in the system size,
$L^{-\beta/\nu}$. This convergence to the critical density is indeed
visible in the insets of Figs.~\ref{fig:den_n0}(a) and
\ref{fig:eta_n0}(a), although we have not attempted to verify the
anticipated scaling behaviour explicitly.

With regard to the behavior of the coexistence number densities as a
function of parent density $n\p$ away from criticality
(Fig.~\ref{fig:den_n0}), we find that the liquid phase density
varies non-monotonically, while the vapor density decreases as the gas
phase branch of the cloud curve is approached. These trends are also
seen in the corresponding MFE
results, though the non-monotonicity in the liquid density is much
weaker and only just perceptible on the scale of the figure. In the
volume fraction 
representation (Fig.~\ref{fig:eta_n0}), on the other hand, the liquid volume
fraction increases monotonically as the gas cloud curve is approached,
while the vapor volume fraction decreases.  Thus the effect of
isothermally reducing $n\p$ from its critical value is analogous to that
seen on lowering the temperature isochorically in a monodisperse system:
the difference in the volume fraction of the two coexisting phases
increases.

\begin{figure}[h]
\includegraphics[type=eps,ext=.eps,read=.eps,width=8.0cm,clip=true]{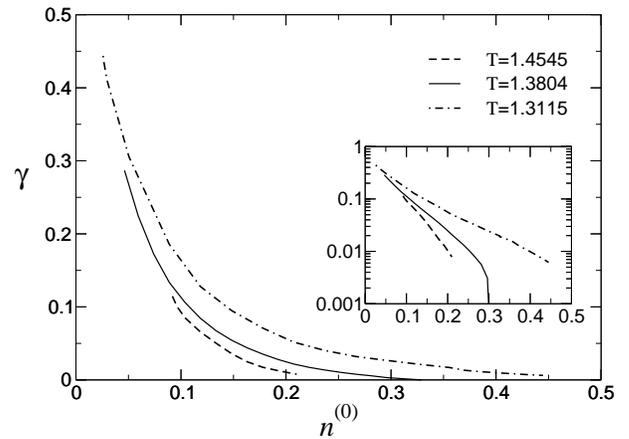}
\caption{Surface tension $\gamma$ for three isotherms at $T>T_c$, $T=T_c$ and
$T<T_c$ and size cutoff $\sigma_c=1.6$; the inset shows the same data
  on a logarithmic scale to emphasize the behaviour near the
  respective liquid (high density) cloud points.}
\label{fig:surfaceten}
\end{figure}

This aspect of the phase behaviour is also manifest in the $n\p$ dependence of
the surface tension, which may be estimated from the form of $p(n)$ via
\cite{BINDER82}
\be
\gamma=\lim_{L\to\infty}\frac{1}{2L^{d-1}}\ln\left[\frac{p^{\rm
      max}(n)}{p^{\rm
      min}(n)}\right]
\label{eq:surfaceten}
  \ee
where $p^{\rm max}(n)/p^{\rm min}(n)$ is the ratio of the peak to trough
probabilities of the order parameter distribution, which provides a
measure of the free energy cost of a planar interface.  Formally
Eq.~(\ref{eq:surfaceten}) is valid only in the limit of sufficiently
large system sizes, where the {\em pair} of liquid-vapor interfaces 
(whose presence is mandated by the periodic boundary conditions) are
effectively non-interacting.  Although we have been unable to study
systems sufficiently large to fully approach this limit in the vicinity
of the critical point, we do regard our results for $\gamma$ as
representative of the bulk for parent densities $n\p<0.29$.
Fig.~\ref{fig:surfaceten} shows our estimates of the surface tension for three isotherms
having temperatures $T<T_c$, $T=T_c$ and $T>T_c$.  One sees that in each
instance the surface tension starts at a large finite value on the gas branch
of the cloud curve, but rapidly decreases with increasing parent
density. On approaching the high density cloud point it falls to zero
for $T=T_c$ as expected for critical phase coexistence, while for
$T>T_c$ and $T<T_c$ it decreases to a finite value as is more clearly
visible in the inset of Fig.~\ref{fig:surfaceten}.

\begin{figure}[h]
\includegraphics[type=eps,ext=.eps,read=.eps,width=8.0cm,clip=true]{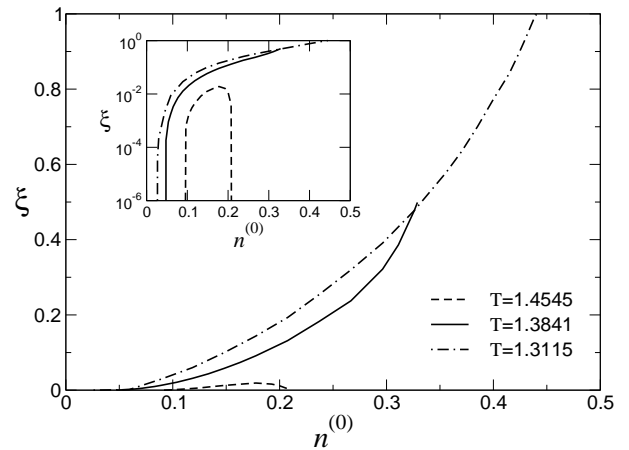}
\caption{MC results for the fractional volume of the liquid phase
  $\xi$ as a function of parent density $n\p$ across the coexistence region,
for three isotherms $T>T_c$, $T=T_c$ and $T<T_c$ and size cutoff
$\sigma_c=1.6$. Inset: Same data on a logarithmic scale.}
\label{fig:xi_isotherm}
\end{figure}

Finally in this section, we point out qualitative differences in the
nature of the coexistence behaviour above and below the critical
temperature. Fig.~\ref{fig:xi_isotherm} plots our simulation
estimates of $\xi$, the fractional volume of the liquid [cf.\ 
Eq.~(\ref{eq:lever})], as a function of $n\p$ for the same three
isotherms as considered above. For the case $T<T_c$, the fractional volume
of the liquid is zero at the gas phase cloud, and steadily increases
with $n\p$ -- in a non-linear fashion -- to reach $\xi=1$ at the liquid
phase cloud point. By contrast for $T=T_c$, $\xi$ increases from zero at
the gas phase cloud, but reaches a maximum of $\xi=0.5$ at the critical
point. Finally for $T>T_c$, $\xi$ starts from zero, increases to a
maximum, and then decreases to zero again as the high density branch of
the cloud curve is reached. Thus the distinguishing feature of the
``supercritical'' (in temperature, i.e.\ $T\geq T_c$) coexistence region
is that the fractional 
volume of the liquid phase never reaches unity. The corresponding MFE
predictions show the same qualitative behaviour (data not shown).

\section{Phase behaviour: cutoff effects}
\label{sec:cutoff}

\begin{figure}[h]
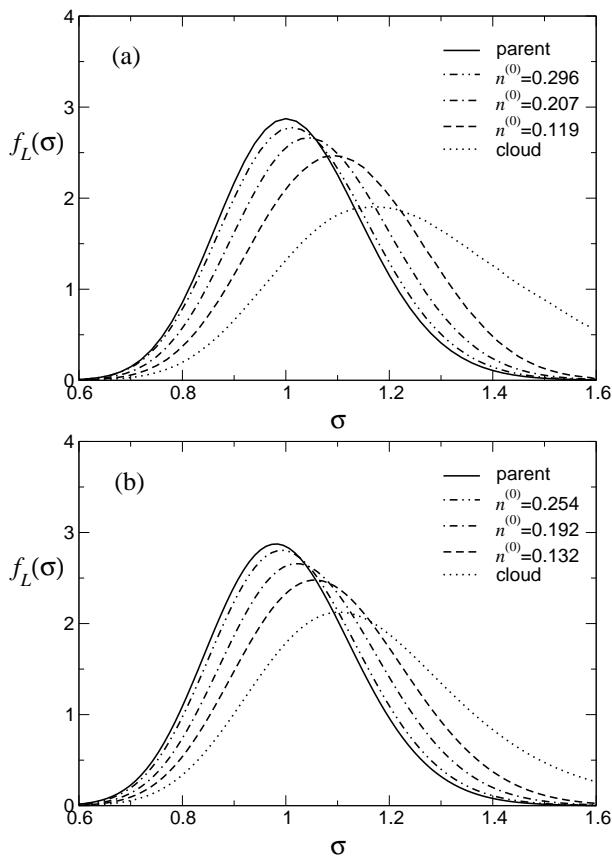

\includegraphics[type=eps,ext=.eps,read=.eps,width=8.0cm,clip=true]{Figs/Fig09a}
\includegraphics[type=eps,ext=.eps,read=.eps,width=8.0cm,clip=true]{Figs/Fig09b}
\caption{
Liquid phase size distributions $f_L(\sigma)$ for a
selection of values of $n\p$ spanning the coexistence region at $T=T_c$.
Data are shown for cutoff
%s $\sigma_c=1.4$ and 
$\sigma_c=1.6$.
{\bf (a)} MC results. {\bf (b)} MFE predictions; densities were chosen to
be at the same relative positions across the coexistence region as for
(a).}
\label{fig:daughter}
\end{figure}

We return now to address the striking feature evident from
Figs.~\ref{fig:cl_sh_n} and \ref{fig:cl_sh_eta} of a large
difference in the measured cloud and shadow curves for the two cutoffs
$\sigma_c=1.4$ and $1.6$. Specifically, the gas phase cloud curve for
$\sigma_c=1.6$ is shifted to much lower densities compared to that for
$\sigma_c=1.4$. This occurs even though both values of $\sigma_c$ are
far in the tail of the parent distribution [cf.\
 Fig.~\ref{fig:schulz50}]. The origin of this effect is to be found
in the character 
of the particle size distributions in the liquid daughter phase.
Fig.~\ref{fig:daughter} shows the form of this distribution for selected
values of $n\p$ on the critical isotherm for 
%
%$\sigma_c=1.4$ and
%
$\sigma_c=1.6$. One observes that as the gas cloud point density is
approached from above, there is a progressive accumulation of weight in
the large-$\sigma$ region of the distribution. Thus despite the fact
that particle sizes around $\sigma_c$ are very rare in the parent, they
occur (by virtue of fractionation) in much higher concentrations in the
liquid. The physical basis for this is the stronger interaction between
the larger particles [cf.\ Eq.~(\ref{eq:uij})]; an enhancement in the
concentration of such particles then yields a substantial free energy
gain at the shorter interparticle separations of the liquid. 

\begin{figure}[h]
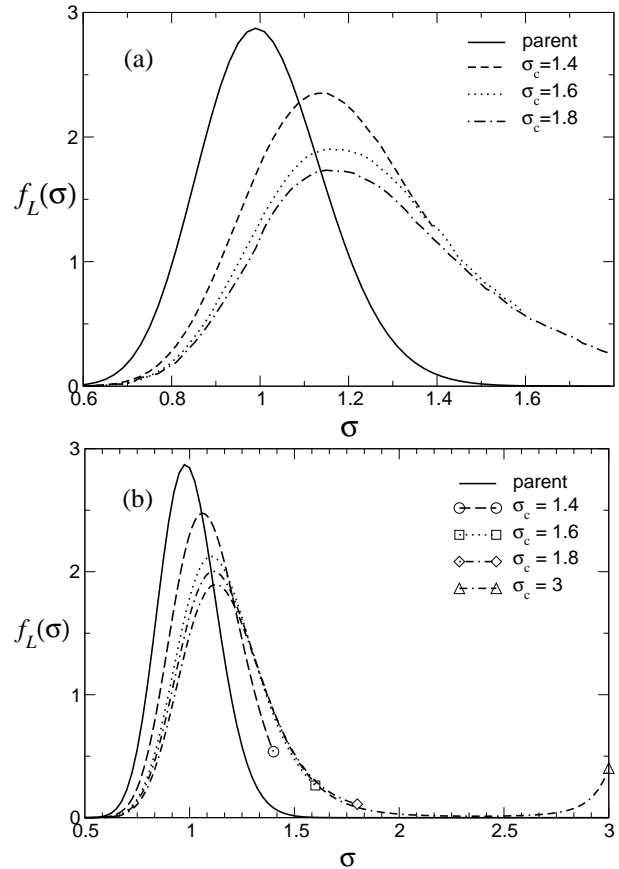

\includegraphics[type=eps,ext=.eps,read=.eps,width=8.0cm,clip=true]{Figs/Fig10a}
\includegraphics[type=eps,ext=.eps,read=.eps,width=8.0cm,clip=true]{Figs/Fig10b}
\caption{{\bf (a)} Size distributions $f_L(\sigma)$ in the liquid shadow
phase at $T=T_c$ for $\sigma_c=1.4, 1.6, 1.8$, together with
the parent distribution $f\p(\sigma)$ plotted for comparison. {\bf(b)}
MFE predictions, including the larger cutoff $\sigma_c=3$.}
\label{fig:shadowdists}
\end{figure}

Clearly, therefore, the choice of cutoff has a profound effect on the
liquid daughter phase distribution, particularly in the low density
region close to the cloud point where the fractionation-induced
enhancement of the large-$\sigma$ tail of the size distribution in the
liquid is
greatest. The magnitude of the truncation effect {\em at} the gas phase
cloud point, for $T=T_c$, is quantified in
Fig.~\ref{fig:shadowdists}(a), which compares the measured forms of the
shadow phase size distribution at cutoffs $\sigma_c=1.4, \sigma_c=1.6$ and
$\sigma_c=1.8$. For these cutoff values, we find that the density at
$\sigma=\sigma_c$ is enhanced compared to the corresponding parent
density by factors of $11.9(1)$, $175(7)$ and $5280(50)$ respectively.
Also shown, in Fig.~\ref{fig:shadowdists}(b), are the corresponding MFE
results for the same choice of cutoffs, together with a further
distribution for the larger cutoff $\sigma_c=3$; we discuss the form
of the latter below.

Given that the liquid shadow phase distribution is highly sensitive to
the cutoff {\em and} that this phase {\em coexists} with the gas cloud
phase, the origin of the sensitivity of the locus of the cloud curve to
the choice of cutoff [as seen in Figs.~\ref{fig:cl_sh_n} and
\ref{fig:cl_sh_eta}] becomes clear. We can now also rationalize the
observation that such significant cutoff effects are restricted to
parent densities {\em below} the critical density. For higher densities,
the shadow phase is a gas of {\em lower} density than the parent. In
this, the concentration of large particles is {\em suppressed} and that
of small particles negligibly enhanced because of their weak
interactions. The shadow size distributions are therefore concentrated
well within the range $0.5\ldots \sigma_c$ so that no noticeable cutoff
dependence arises. 

\begin{figure}[h]
\includegraphics[type=eps,ext=.eps,read=.eps,width=8.0cm,clip=true]{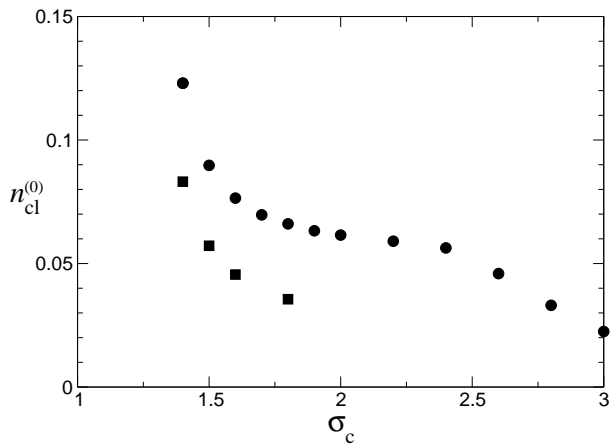}
\caption{The variation of
the gas cloud point density $n\p\cl$ at $T=T_c$ as a function of
$\sigma_c$, as obtained from MC simulations (squares) and MFE
calculations (circles).}
\label{fig:cloudshift}
\end{figure}

\begin{figure}[h]
\includegraphics[type=eps,ext=.eps,read=.eps,width=8.0cm,clip=true]{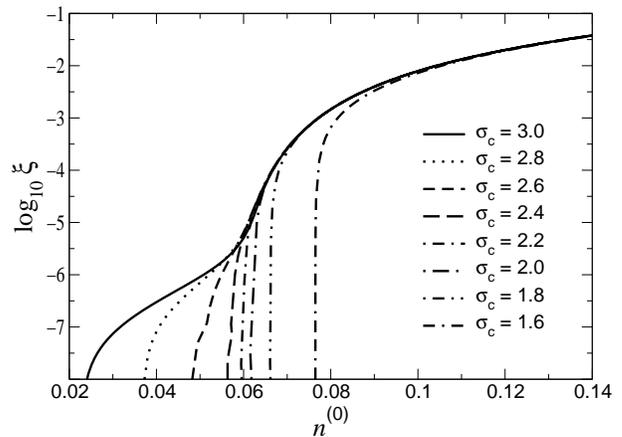}
\caption{Plot of the fractional volume of the liquid phase versus
  parent density as
  predicted by the MFE theory at the critical temperature
  $t=t_c=0.437$. Results are displayed for a selection of size cutoffs as
  shown.}
\label{fig:xi_shift}
\end{figure}

The observed decrease in the gas cloud point density $n\cl\p$ with increasing
$\sigma_c$ prompts the question as to whether the gas phase cloud point
density would eventually tend to a zero or nonzero limit as $\sigma_c$
is increased. Fig.~\ref{fig:cloudshift} shows the simulation results and
MFE predictions on the critical isotherm. The former exhibit a further
strong decrease of $n\p\cl$ by $\approx 25\%$ between $\sigma_c=1.6$ and
$1.8$; the latter suggest that this trend continues and that the cloud
point density tends to zero for large $\sigma_c$. Such an unusual effect
has previously been seen in theoretical studies of polydisperse hard
rods with broad length distributions~\cite{SpeSol03a} and in a
Flory-Huggins model for polymers \cite{SOLC}. It is also predicted to
occur in solid-solid phase separation of polydisperse hard spheres
\cite{HS_longpaper}, though only for large $\sigma_c$ and distributions
with fatter than exponential tails. In the present model, however, the
decrease of $n\p\cl$ is clear even for $\sigma_c$ of $O(1)$, i.e.\ of
the same order as $\bar{\sigma}$. The physical origin of the decrease of
$n\p\cl$ to zero is the appearance (for large $\sigma_c$) of a second
peak in the shadow phase size distribution near $\sigma_c$
(Fig.~\ref{fig:shadowdists}(b)). As with hard rods, we expect this
second peak to eventually dominate as $\sigma_c$ increases so that the
shadow phase comprises ever more strongly interacting particles whose
sizes are concentrated near $\sigma_c$. The appearance of the second
peak in the size distribution also correlates directly with a
significant extension of the coexistence region towards smaller parent
densities, as shown in Fig.~\ref{fig:xi_shift}. This extension shifts
the cloud point to lower densities, leaving in its wake a region where
the fractional volume occupied by the liquid phase is extremely small,
below $10^{-6}$. This region then crosses over at parent densities around
$n\p=0.06$ to more conventional phase behaviour, where $\xi$
eventually becomes cutoff-independent. These qualitative
features are similar to those observed to hard rods with broad length
distributions~\cite{SpeSol03a}.

It is natural to enquire as to the conditions under which cutoff
dependences can be expected be occur. The answer will, in general,
depend on both the choice of the form of the parent size distribution
$f\p(\sig)$ and the 
size-dependence of the strength of the interaction between particles. It is
straightforward to show~\cite{SOLLICH02} that the density distribution
of the liquid shadow phase is related to the parent via:
\bea
\rho^{\rm sh}(\sig)&=&\rho\p(\sigma) \exp\left({-\beta[\mu^{\rm sh}_{\rm
	ex}(\sig)-\mu^{\rm cl}_{\rm ex}(\sig)]}\right)\\
               &\approx& \rho\p(\sigma) \exp\left[{-\beta\mu^{\rm sh}_{\rm
      ex}(\sig)}\right]
\label{ratio}
\eea
when the cloud point density is small. Thus when the excess activity
$\exp[\beta\mu^{\rm sh}_{\rm ex}(\sigma)]$ in
the shadow phase decreases faster with $\sigma$ than the
corresponding decay in the parent size distribution, the shadow
density distribution will be peaked at the cutoff.

We now analyse this scenario in more detail, for a more general
size-dependence of the interaction strength,
$\epsilon_{ij}=(\sigma_i\sigma_j)^\alpha$. Our hypothesis is that, for
suitable parent distributions $f\p(\sigma)$, the cloud point will move
to vanishing density as the cutoff $\sigma_c$ is made large, while the
corresponding shadow phase will consist only of particles with sizes
close to $\sigma_c$. The shadow is then a scaled version of a
``standard'' (monodisperse, unit particle diameter) LJ system, but
effectively at a temperature $T/\sigma_c^{2\alpha}$ because all
interactions are stronger by a factor $(\sigma\sigma')^\alpha\approx
\sigma_c^{2\alpha}$; this effective 
temperature decreases to zero as $\sigma_c$ is made large. To coexist
with the (infinitely, in the limit $\sigma_c\to 0$) dilute cloud phase,
the shadow also has to be at zero pressure. At these zero temperature,
zero pressure conditions, the shadow phase will be a crystalline solid,
more precisely the ground state of the effectively monodisperse LJ
system. To check that this scenario is self-consistent, we now need to
work out the excess chemical potential $\mu_{\rm ex}^{\rm sh}(\sigma)$
of such a solid. It is convenient to use the Widom insertion expression
$\exp[-\beta\mu_{\rm ex}^{\rm sh}(\sigma)]=\langle \exp(-\beta
u_\sigma)\rangle$; the  average is over the position where the test
particle of diameter $\sigma$ is inserted, uniformly over the whole
system, and $u_\sigma$ is the interaction potential of this particle
with all others. Given the assumed size-dependence of the interaction
strength, we have $u_\sigma=(\sigma\sigma_c)^\alpha u_1$, where $u_1$ is
the interaction potential of a unit diameter test particle inserted into
a standard LJ solid. Thus $\exp[-\beta\mu_{\rm ex}^{\rm
sh}(\sigma)]=\langle \exp[-\beta(\sigma\sigma_c)^\alpha u_1]\rangle$.
When $\sigma_c$ is large, the average will be dominated by the minimum
value $u_{1,\rm min}<0$ of $u_1$, i.e.\ the optimal insertion positions.
Ignoring subexponential corrections then gives $\mu_{\rm ex}^{\rm
sh}(\sigma)=-(\sigma\sigma_c)^\alpha |u_{1,\rm min}|$. The excess
chemical potential at $\sigma=\sigma_c$ thus scales as
$\sigma_c^{2\alpha}$; looking at Eq.~(\ref{ratio}), the hypothesized
cutoff-dominated shadow phase can then exist for parent size
distributions decaying more slowly than $f\p(\sigma)\sim \exp(-{\rm
const}\times\sigma^{2\alpha})$. Toward smaller $\sigma$, $-\mu_{\rm
ex}^{\rm sh}(\sigma)\sim (\sigma \sigma_c)^{2\alpha}$ decreases by an
amount of $O(1)$ already for $\sigma_c-\sigma \sim \sigma_c^{1-2\alpha}
\ll \sigma_c$. This fast decrease shows that also the assumption of a
shadow size distribution  sharply peaked towards $\sigma_c$ is
self-consistent. For the particular case $\alpha=1$ studied in the rest
of this paper, we have that the limiting parent size distribution where
cutoff effects start to appear is Gaussian. The Schulz distribution has
a substantially fatter (exponential) tail and it is therefore clear that
cutoff dependences should arise, as observed.

As we have seen, the MFE results (cf.\ Fig.~\ref{fig:shadowdists})
predict that, as $\sigma_c$ is  increased, so the shadow phase comprises
ever more strongly interacting particles whose sizes are concentrated
near $\sigma_c$. For large enough cutoffs, the general arguments
advanced above show that this shadow phase must be a solid because it is
 effectively at very low temperature (as well as low pressure). Since
the simple approximate free energy used for our theoretical predictions
does not include a branch that could describe such solid phases, we have
attempted to investigate this possibility via simulation. Unfortunately,
owing to the  computational burden of simulating systems with large
cutoffs, we could not access the coexistence regime directly. We were,
however, able to study the regime of metastable coexistence lying
between the cloud curve and a system-size dependent ``effective
spinodal'' \cite{BINDER03}. For large cutoffs, this region occurs at
very low $n\p$, and accordingly one can, to a good approximation, assign
the requisite chemical potential distribution on the basis of the value
of the second virial coefficient, as described in
Appendix~\ref{sec:append}. 

\begin{figure}[h]
\includegraphics[type=eps,ext=.eps,read=.eps,width=8.0cm,clip=true]{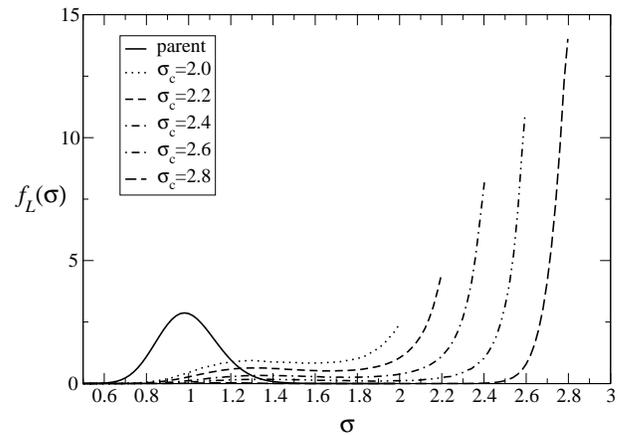}
\caption{Size distribution $f_L(\sigma)$ in the metastable
liquid phase for cutoffs in the
range $\sigma_c=2.0$ to $\sigma_c=2.8$. Also shown for comparison is the
parent form $f\p(\sigma)$.}
\label{fig:metastable}
\end{figure}

\begin{figure}[h]
\includegraphics[type=eps,ext=.eps,read=.eps,width=6.0cm,clip=true]{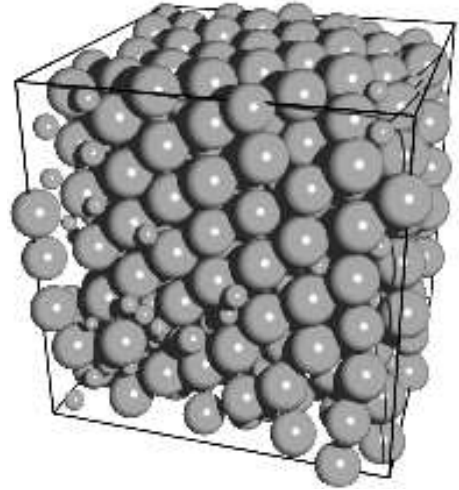}
\caption{Snapshot configuration of the quasi-monodisperse metastable solid
which coexists with the gas phase at $n\p=1\times 10^{-7}$ for $\sigma_c=2.8$.}
\label{fig:freeze}
\end{figure}

Using this approach, we have studied the form of the size distribution
in the metastable liquid
daughter phase for a range of large cutoff values.
The results in Fig.~\ref{fig:metastable} show that,
in accord with the MFE predictions, as $\sigma_c$ increases there is
an accumulation of weight at $\sigma=\sigma_c$. Interestingly
too, we find that for $\sigma_c=2.8$ the liquid spontaneously freezes to
an f.c.c.\ crystal structure, as shown in Fig.~\ref{fig:freeze}. We
emphasize that this occurs for $n\p$ values close to the effective
spinodal point, not at the cloud point itself. However, given the
above theoretical considerations it seems likely
that, for $\sigma_c$ values larger than those presently accessible to
direct simulation at the cloud point, the freezing will occur from
the stable liquid phase~\cite{SEAR}. 

\begin{figure}[h]
\includegraphics[type=eps,ext=.eps,read=.eps,width=7.0cm,clip=true]{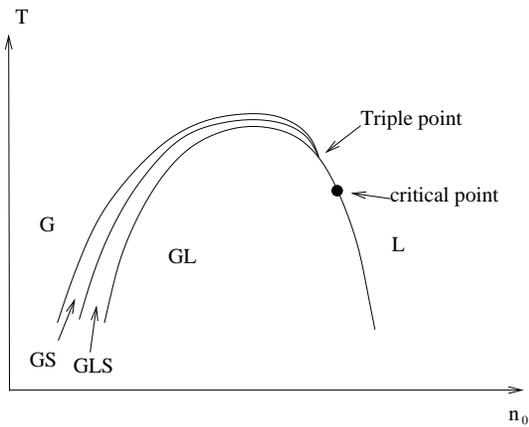}
\caption{Schematic representation of a possible phase diagram for
  large size cutoffs $\sigma_c$. The letters indicate the nature of
  the phases (G: gas, L: liquid, S: solid).}
\label{fig:posspd}
\end{figure}

In view of the likelihood of gas-solid coexistence at large $\sigma_c$
and small $n\p$, one can speculate as to the character of the overall phase
diagram in this regime. One possibility is that depicted schematically
in Fig.~\ref{fig:posspd}. Here on increasing $n\p$ from the stable gas
region, at low $T$, the system reaches a gas-solid (GS) cloud point at
which the gas splits off an infinitesimally small quantity of
quasi-monodisperse solid. However, as more of this solid is
formed it must become more polydisperse in order for the system
overall to preserve the parent size distribution.
Hence on further increasing $n\p$, the increasing
polydispersity of the solid forces it to split off a liquid phase and
the system must enter a region of gas-liquid-solid (GLS) coexistence.
Finally, once all the solid has melted, there is a regime of
gas-liquid (GL) coexistence,  as is also seen for smaller cutoffs.
It seems likely that at higher temperature the solid phase
would not be stable, leading to a gas-liquid-solid triple
point as shown.

\section{Conclusions}
\label{sec:concs}

In summary, we have deployed state-of-the-art MC simulation techniques
and MFE calculations to study in detail the phase behaviour of a model
fluid in which polydispersity affects both the particle sizes and the
strength of their interactions. The latter aspect in particular is
primarily responsible for a dramatic separation of cloud and
shadow curves compared to a previous study of the purely size-disperse
case \cite{WILDING04}. We have determined the cloud and shadow curves
for our model, as well as the phase behaviour along certain dilution
lines which span the coexistence region. We find that the locus of the
cloud curve is acutely sensitive to the choice of the upper cutoff on the
parent particle size distribution, even when this cutoff lies far in the
tail of the distribution. Such effects imply that in experiments on
polydisperse fluids (see e.g.~\cite{ERNE05}) it may be important to
monitor and control carefully the tails of the size (or charge, etc)
distribution. Otherwise undetected differences could lead to large
sample-to-sample fluctuations in the observed phase behaviour. 

The origin of the observed cutoff dependences has been traced to
extremely pronounced fractionation effects, the nature of which we have
elucidated in terms of the character of the size distribution of the
shadow phase. We have also provided criteria for determining which
combination of parent size distributions and interaction potentials can
be expected to display cutoff effects. Our theoretical considerations,
together with additional simulation evidence, suggest that, in the limit
of very large cutoffs, the cloud point density tends to zero and new
phases appear, such as a region of coexistence between a gas and a
quasi-monodisperse crystal.

Finally, with regard to outstanding questions that have not been
considered in the present work, we would highlight the nature of the
critical behaviour. Specifically, nothing is known concerning the
existence and nature of singularities in the cloud and shadow curves in
the critical region, or indeed what constitutes a suitable choice of
order parameter for characterizing the critical behavior.  Intriguing
experimental results on polydisperse polymers \cite{KITA} suggest that
the critical exponents are Fisher renormalized, though the reasons for
this appear unclear at present. We hope to investigate these issues in
future work.

\appendix 
\renewcommand{\theequation}{\Alph{section}\arabic{equation}}

\section{Virial estimate of chemical potential distribution at low parent density}

\label{sec:append}

For a monodisperse system, the chemical potential can be
written to first order in the density as

\begin{equation}
\beta\mu_{\rm ex}=2\rho\sigma^3 B_2^\star(T)
\end{equation}
with $B_2(T)$ the reduced second virial coefficient \cite{HANSEN}.

For our polydisperse LJ potential, Eq.~(\ref{eq:uij}), this generalizes to
\begin{equation}
\beta\mu_{\rm ex}(\sigma)=2\int_0^{\sigma_c}\!\!
d\sigma^\prime\,\rho(\sigma^\prime)
\left(\frac{\sigma+\sigma^\prime}{2}\right)^3
B_2^\star(T/\sigma\sigma^\prime)
\label{eq:mu_ex}
\end{equation}
with
\begin{equation*}
B_2^\star=-2\pi  \int_0^\infty\!\!dr\,
r^2\exp\left\{\frac{-4\sigma\sigma^\prime}{T}\left[\left(\frac{\bar \sigma}{r}\right)^{12}-\left(\frac{\bar \sigma}{r}\right)^{6}\right]-1\right\}
\end{equation*}

Numerical evaluation of the double integral in (\ref{eq:mu_ex}) is
simplified by noting that for the (untruncated) Lennard-Jones potential,
$B_2^\star(T)$ can be expressed as a series:

\begin{equation}
B_2^\star(T)=\frac{-2\pi}{3}\sum_{n=0}^\infty \frac{1}{4n!}\left(\frac{4}{T}\right)^{(2n+1)/4}\Gamma\left(\frac{2n-1}{4}\right)
\end{equation}
with $\Gamma(z)$ the complete Gamma function. We have not corrected for the
presence of a potential truncation.

\acknowledgments
The authors acknowledge support of the EPSRC, grant numbers
GR/S59208/01 and GR/R52121/01. PS acknowledges the warm
hospitality of the Isaac Newton Institute, Cambridge, where part of
this work was completed.

\end{document}